\documentclass[a4paper,fleqn]{cas-dc}

 \usepackage[numbers]{natbib}

\def\tsc#1{\csdef{#1}{\textsc{\lowercase{#1}}\xspace}}
\tsc{WGM}
\tsc{QE}
\tsc{EP}
\tsc{PMS}
\tsc{BEC}
\tsc{DE}

\begin{document}
 \shorttitle{VRT for Photon Transport Simulation in LXe Detectors}
 \shortauthors{A.P. Colijn et~al.}

\title [mode = title]{Monte Carlo Simulation Variance Reduction Techniques for Photon Transport in Liquid Xenon Detectors}                      

%

\author[1]{S. Bruenner}

\address[1]{Nikhef  and  the  University  of  Amsterdam,  Science  Park 105,  1098 XG  Amsterdam, The Netherlands}

\author[1,2]{A.P. Colijn}
\cormark[1]
\ead{a.p.colijn@uva.nl}
\address[2]{Institute for Subatomic Physics, Utrecht University, Princetonplein 1, 3584 CC Utrecht, The Netherlands}

\author[1]{M.P. Decowski}

\author[1]{O.V. Kesber}

\cortext[1]{Corresponding author}

\begin{abstract}
Monte Carlo simulations are a crucial tool for the analysis and prediction of various background components in liquid xenon (LXe) detectors. With improving shielding in new experiments, the simulation of external backgrounds, such as induced by gamma rays from detector materials, gets more computationally expensive. We introduce and validate an accelerated Monte Carlo simulation technique for photon transport in liquid xenon detectors. The method simulates photon-induced interactions within a defined geometry and energy range with high statistics while interactions outside of the region of interest are not simulated directly but are taken into account by means of probability weights. For a simulation of gamma induced backgrounds in an exemplary detector geometry we achieve a three orders of magnitude acceleration compared to a standard simulation of a current ton-scale LXe dark matter experiment.
\end{abstract}

%


\maketitle

\section{Introduction}
In the last decades, liquid xenon (LXe) detectors became a leading technology for rare event searches in astro particle physics. Many of the most stringent limits for the direct detection of dark matter \cite{Aprile:2018final,LZ:2018} or on the neutrinoless double-beta decay \cite{EXO-200:2019} have been set by LXe experiments. One major advantage of LXe is its high mass density and the resulting high self-shielding capability against external radiation. Background induced by gamma radiation from radioactive decays in detector materials is efficiently reduced in the central LXe volume due to the shielding of the outer xenon layers. In order to optimize the detector sensitivity, a precise knowledge of the remaining background components is essential. Therefore, many experiments do intensive measurements of trace radioactivity impurities in the detector materials \cite{Aprile:2017screening,Akerib:2020com}. This data is used as input for Monte Carlo simulations of the gamma background in the central LXe volume \cite{Aprile:2015uzo,LUX:2015MC}. For future large scale  detectors, employing up to 50\,t of LXe \cite{Aalbers:2016DARWIN}, these simulations will become computationally expensive as a large number of gamma emissions need to be simulated in order to achieve a robust prediction of background events in the center of the detector, particularly in the low energy regime of a few tens of keV. Variance reduction techniques (VRT) such as \textit{importance sampling} \cite{doi:10.1002/wics.56} are one approach to increase the precision of Monte Carlo simulations while at the same time reducing the computational effort.\\
In this paper, we introduce a VRT dedicated to the simulation of external gammas in LXe detectors. The statistical uncertainty is decreased with respect to standard Monte Carlo simulations by means of importance sampling. The simulation can be restricted to photon interactions in the region of interest (ROI) of a rare event signal. The underlying probability density functions (pdf) are thus over-sampled in the ROI, while they are not sampled  in other regions of phase space. The non-sampled events are taken into account by applying corresponding weights to the obtained simulation result in the ROI. Section 2 introduces the VRT used for the following simulations in this paper. For validation, we simulate gamma interactions in LXe by means of the standard Monte Carlo approach and compare those to the VRT simulation results. A discussion of the achieved acceleration due to VRT is presented in section 3. We conclude with a summary of of our results in section 4.

\section{Simulation Method and Validation}\label{sec:method}

To study the variance reduction techniques proposed in this paper, we developed a photon transport Monte Carlo code in Python. It implements both a standard simulation technique and a variance reduction method to validate our acceleration techniques. During the photon transport only the photo-electric effect and Compton scattering are taken into account, with the mass attenuation coefficients $\mu$ as provided by NIST \cite{NIST}. If a photo-electric absorption takes place, all the photon energy is assumed to be deposited locally and further tracking of the photo-electron is not done: this is a reasonable assumption since the photo-electrons will only traverse a distance of a few 100~$\mu$m in liquid xenon, which is below the position resolution of current LXe experiments \cite{Aprile:2019bbb}. In the case of a Compton scattering event, we use the Klein-Nishina differential cross-section to calculate the scattering angle, while taking into account the binding energy of electrons with appropriate form factors \cite{Hubbell:1975formfactors}. These form factors are especially important in the regime of low-energy deposits in the keV range, and thus for dark matter searches. The MC code does not take into account pair creation. Thus, its applicability is in principle limited to gamma ray energies below the pair creation threshold of 1.02~MeV. However, it is safe to assume that the actual results are reliable up to energies of $\sim 1.5$~MeV since the pair-production cross-section is about one order of magnitude below the incoherent cross-section.

\begin{figure}[htb]
  \centering
    \includegraphics[width=0.4\textwidth]{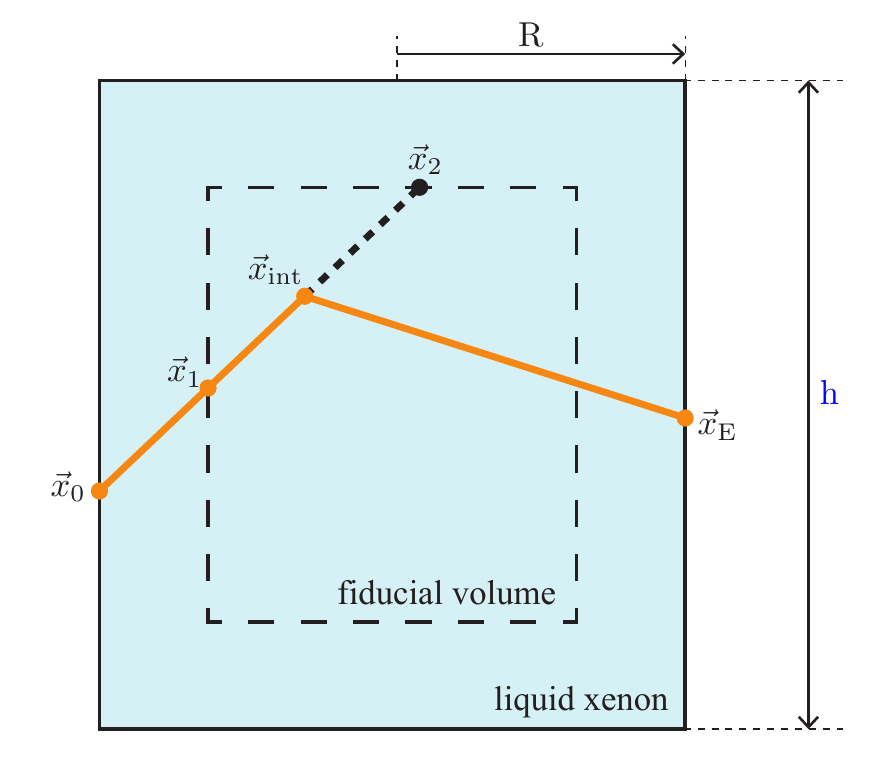}
    \caption{Geometry for the simulation studies presented in this paper, representing a simplified dark matter experiment similar to LZ \cite{Akerib:2019fml} or XENONnT \cite{xenonnt}. Most high-energy gamma rays originate from the boundary of the liquid xenon volume, while signal events occur in the central fiducial volume (FV). The orange line represents a possible trajectory of a gamma ray through the liquid xenon, scattering once.}
    \label{fig:setup}
\end{figure}
We use a cylinder with a radius of $R=65$~cm and a height of $h=150$~cm filled with ~6\,t of LXe as a benchmark model (see Figure \ref{fig:setup}). For simplicity, the detector walls or other detector components were not assumed to be present. Gamma rays can be emitted from random locations on the cylinder's surface. 
The fiducial volume (FV) is a cylinder with a radius of 57~cm and a height of 134~cm at the center of the xenon volume, such that it is surrounded by 8~cm of LXe on all sides. Such a geometry mimics dark matter detector like for example LZ \cite{Akerib:2019fml} or XENONnT \cite{xenonnt}. It has to be noted, however, that given the simplicity of the geometry this study should only be viewed as a feasibility and validation study of the acceleration technique discussed below.

\subsection{Variance Reduction Technique}\label{sec:simulation}
A simulated gamma ray with a certain primary position and direction of propagation has a probability $p_\gamma$ to create an interaction in the pre-defined FV and energy region. Obviously, $p_\gamma$ is dependent on the gamma ray's initial parameters. We factorize the event probability as
\begin{equation}
\centering
p_\gamma = \prod_{i=1}^4 p_i \cdot p_s \quad ,
\label{eq:p_i}
\end{equation}
where the probabilities $p_i$ correspond to well-described processes such as the adsorption probability of a photon when traveling through LXe. We will introduce and discuss those processes, four in total, later in this section. $p_s$, on the other hand, is the combined probability for other processes which are not explicitly taken into account. Instead of determining $p_i$ by means of simulation, as it is done in a standard MC approach, we calculate them for each created gamma on an event-by-event base. As a consequence, the probability for a simulated gamma to induce an interaction in the ROI is increased to $p_\gamma = p_s$ since the processes described by $p_i$ will not terminate the simulated event. Instead they act as a weight on the events simulated in the ROI to retain the correct $p_\gamma$ from equation \ref{eq:p_i}. The probability weights $p_i$ refer to different photon transport or interaction processes and are determined as follows.\\
A gamma ray pointing towards the direction of the FV has a probability $p_1$ to be transported to the edge of the FV, $\vec{x}_1$ in Figure \ref{fig:setup}, without undergoing an interaction. $p_1$ is dependent on the path length $|\vec{x}_1 - \vec{x}_0|$ and is calculated as
\begin{equation}
\label{eq:p1}
    p_1 = \exp{\left(-\rho ~ \mu(E_{\gamma}) \,|\vec{x}_1 - \vec{x}_0|  \right)} \quad ,
\end{equation}
where $\rho$ is the density of LXe $(\sim 3\mbox{g/cm}^3)$ and $\mu(E_{\gamma})$ is the mass attenuation coefficient.
Once the gamma ray reaches the FV, it has a probability $p_2$ to undergo an interaction before leaving the sensitive volume at position $\vec{x}_{2}$. In the accelerated simulation, we force an interaction in the FV and apply the weight
\begin{equation}
    p_2 = 1 - \exp{(- \rho \, \mu(E_{\gamma})\, |\vec{x}_2 - \vec{x}_1|)} \quad.
\end{equation}
Simulations in the low energy region are particularly computationally expensive. We can accelerate the simulation by restricting the interactions to energies smaller than a maximum energy deposit of $E_{max}$. The photo-electric effect is prohibited if the energy of a gamma is larger than $E_{max}$ and for Compton scattering only scattering angles associated with interactions below the maximum energy are allowed. Simulated interactions need to be weighted by the probability for an interaction below $E_{max}$ which is given by
\begin{equation}
    p_3 =\frac{1}{\sigma} \int_{0}^{E_{max}} \frac{d \sigma}{dE}dE \quad ,
\end{equation}
where $\sigma = \sigma_{pe} + \sigma_{cs}$ is the total cross-section of gamma-rays including the photo-electric effect and Compton scattering, respectively. Finally, in order to produce a single scatter event the gamma needs to leave the sensitive volume without any further energy deposition. The corresponding probability is calculated similarly to Eq. \ref{eq:p1} by 
\begin{equation}
    p_4 = \exp{\left(-\rho ~ \mu(E_{\gamma}) \,|\vec{x}_E - \vec{x}_{int}|  \right)} \quad .
\end{equation}
We restricted the discussion so far to an event topology where the gamma leaves the LXe volume after a single interaction. However, the introduced acceleration methods can be easily extended to multiple interaction events as we will discuss later. In that case the probability weights are adapted to a gamma undergoing a certain number of interactions within the pre-defined energy range and then leaving the FV or getting terminated in the last interaction by photo-electric absorption.

\subsection{Validation}\label{sec:validation}
\begin{figure*}
\centering
    \centering
    \includegraphics[width=0.8\textwidth]{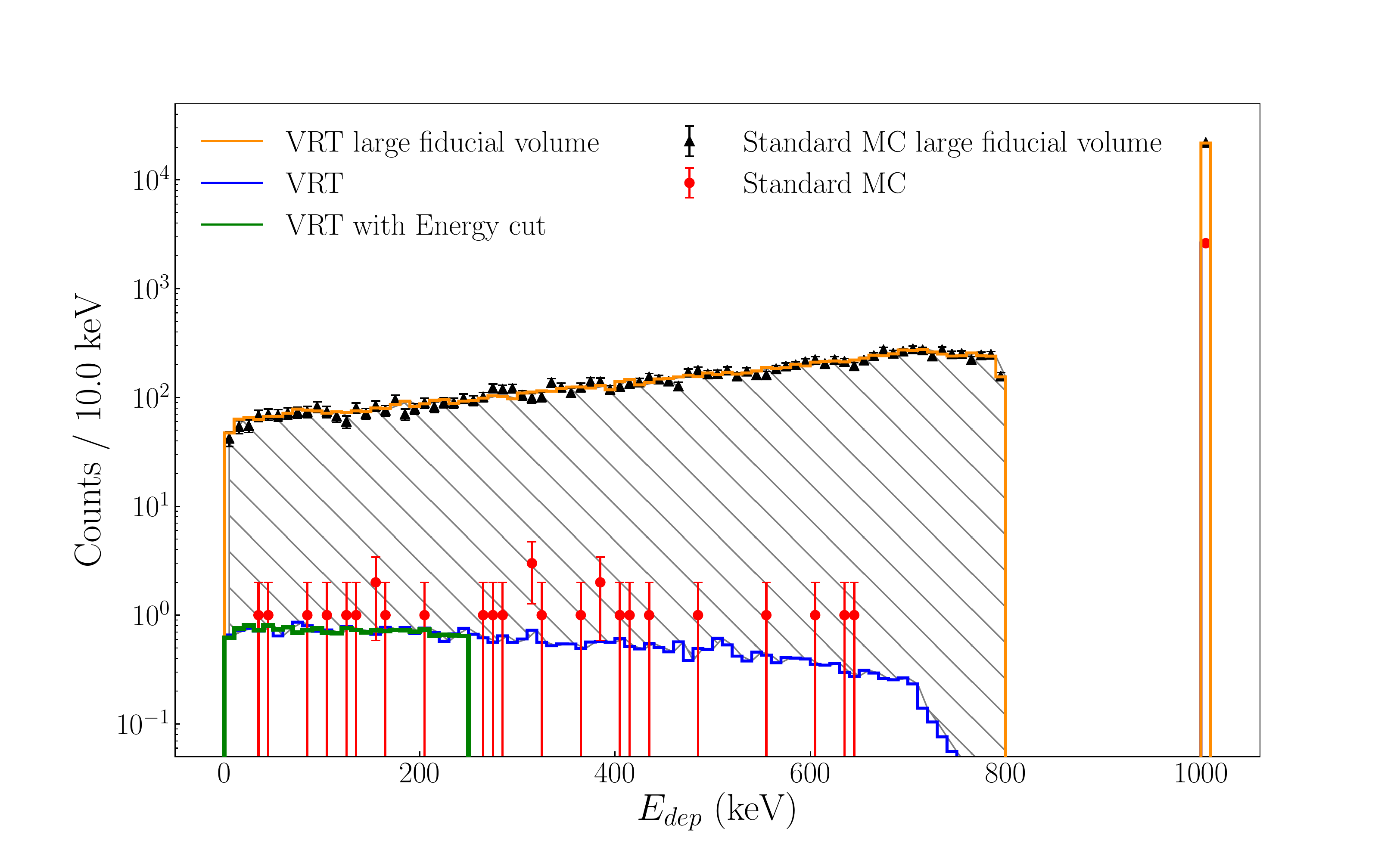}
    \caption{The energy spectra of single scatter events originating of $10^6$ simulated gamma-rays of 1\,MeV energy. Results obtained with the standard MC methods without/with a fiducial volume (FV) are indicated by black/red markers, respectively. The accelerated simulation spectra, indicated by colored histograms, are in agreement with the standard MC but have higher statistics in the FV.}
    \label{fig:val_full_spectrum_large}
\end{figure*}

\begin{figure}
\centering
    \centering
    \includegraphics[width=1.0\linewidth]{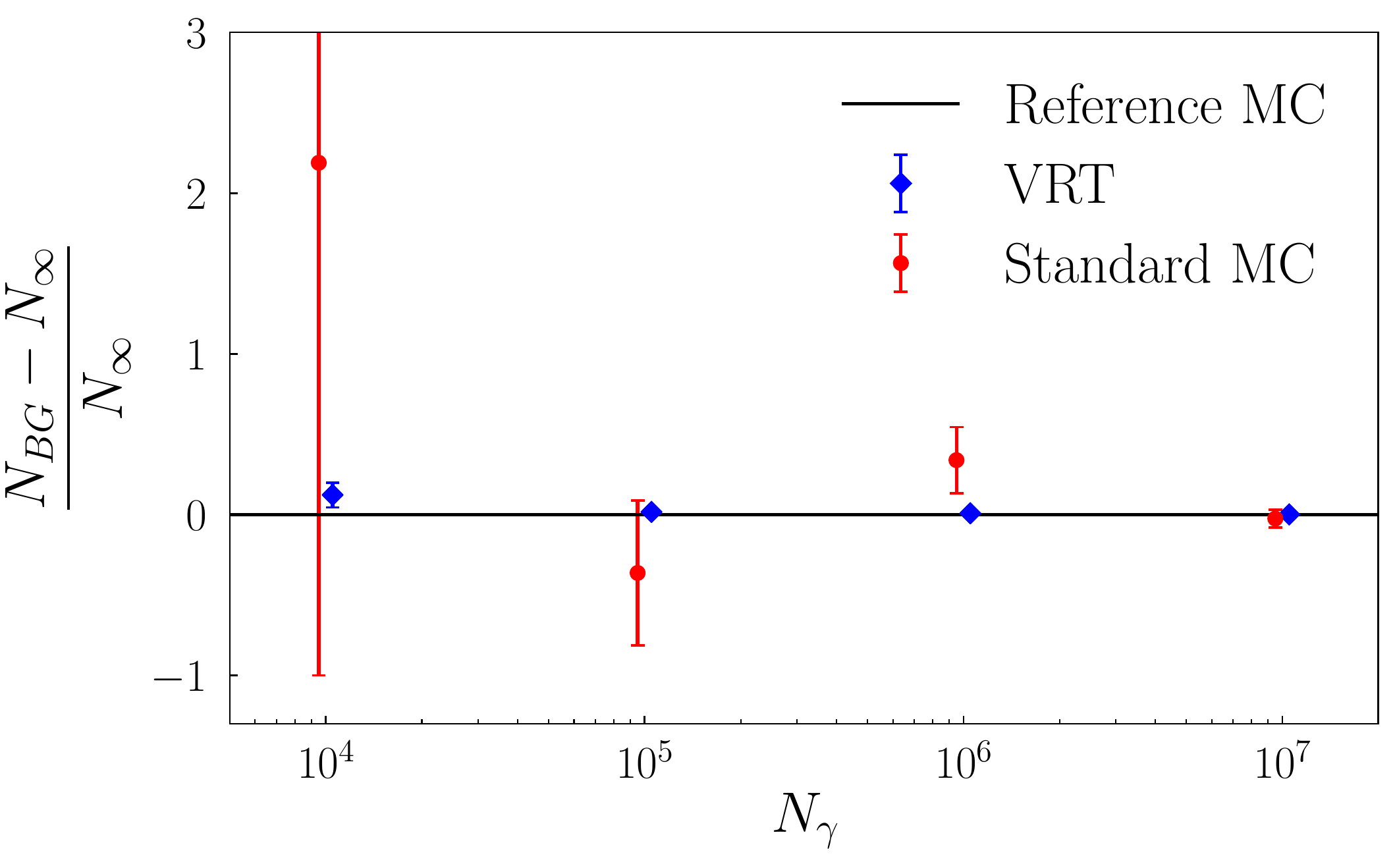}
    \caption{The relative difference between the predicted number of interactions, $N_{BG}$, and the expected number $N_\infty$, obtained in a high statistic accelerated MC. The standard MC prediction (red) converges to $N_\infty$ for a large number of simulated events $N_\gamma$ but slower than the accelerated MC (blue).}
    \label{fig:val_totalweight}
\end{figure}

In order to validate our accelerated simulation method, we compare the results with those obtained from a standard MC simulation. To simplify the discussion, we use the simulation of single scatter events induced by mono-energetic gamma rays of $1\,$MeV as an example. In the first validation step we did not define a FV. This means the weights for transportation until the FV and for the maximum energy deposition are unity, i.e., $p_1 = p_3 = 1$ . The obtained energy spectra of the standard MC method and the accelerated simulation are shown in Figure \ref{fig:val_full_spectrum_large}.
Both show excellent agreement with each other in their prediction of the intensity of the photo-peak at $1\,$MeV and the Compton spectrum. This proves that the physics of gamma interactions is implemented in a consistent way among the simulation methods. In the second step we compare the energy spectra of single scatter interactions within a FV which is shielded by 8\,cm of LXe on all sides. The calculated weight $p_1$ is used in the accelerated simulation. The spectrum of the standard MC suffers from very low statistics apart of the photo-absorption peak (see Figure \ref{fig:val_full_spectrum_large}). The accelerated MC, in contrast, obtains a clear spectrum with high statistics since no initial gamma gets terminated on its way to the FV. Both spectra agree with each other after applying all weights to the accelerated MC. The decreasing shape of the Compton spectrum is explained by the single scatter requirement and reflects the decreasing probability of a gamma to escape the detector without any further interactions as a function of the gamma's first energy deposition. The third test is an accelerated MC simulation with a maximum energy requirement of $E_{max}=250\,$keV. The fact that it is in agreement with the previous simulation without an energy cut-off is a validation of the weight $p_3$ in the accelerated simulation.\\
As a figure of merit, we can predict the number of gamma induced background events $N_{BG}$ in the FV using the standard MC and the accelerated MC approach with increasing statistics. We use an accelerated MC simulation of $10^9$ events and its expectation value $\mu_{BG}$ as a reference simulation. Figure \ref{fig:val_totalweight} shows that the standard MC expectation value of events within the FV (no energy cut applied) converges to the reference value with increasing number of simulated gammas $N_\gamma$.
The error on the standard MC is obtained by $\sqrt{N_{BG}}/\mu_{BG}$ since the error on the reference MC is negligible. For comparison we also show the accelerated MC simulation (blue marker) which converges much faster to the reference value. The error on the accelerated MC is determined from the variance of 10 simulations as we will discuss below.\\
Both validation studies, on the spectral shape as well as on the absolute number of simulated background events in a FV, show good agreement between the standard MC and the accelerated simulation.

\section{Acceleration factor} \label{sec:Acceleration}
\begin{figure*}
        \centering
            \includegraphics[width=0.48\textwidth]{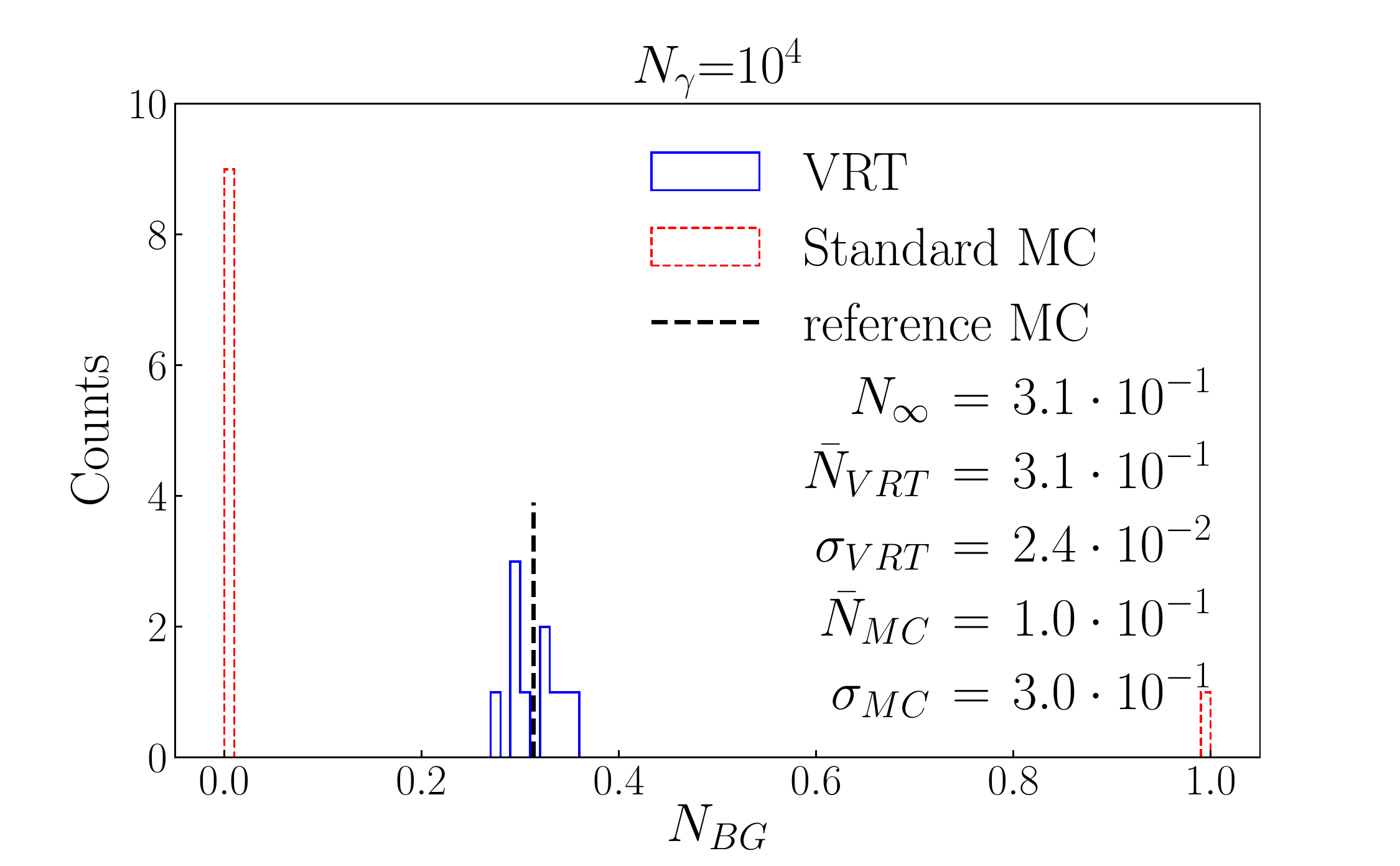}
        \hspace*{\fill}
            \includegraphics[width=0.48\textwidth]{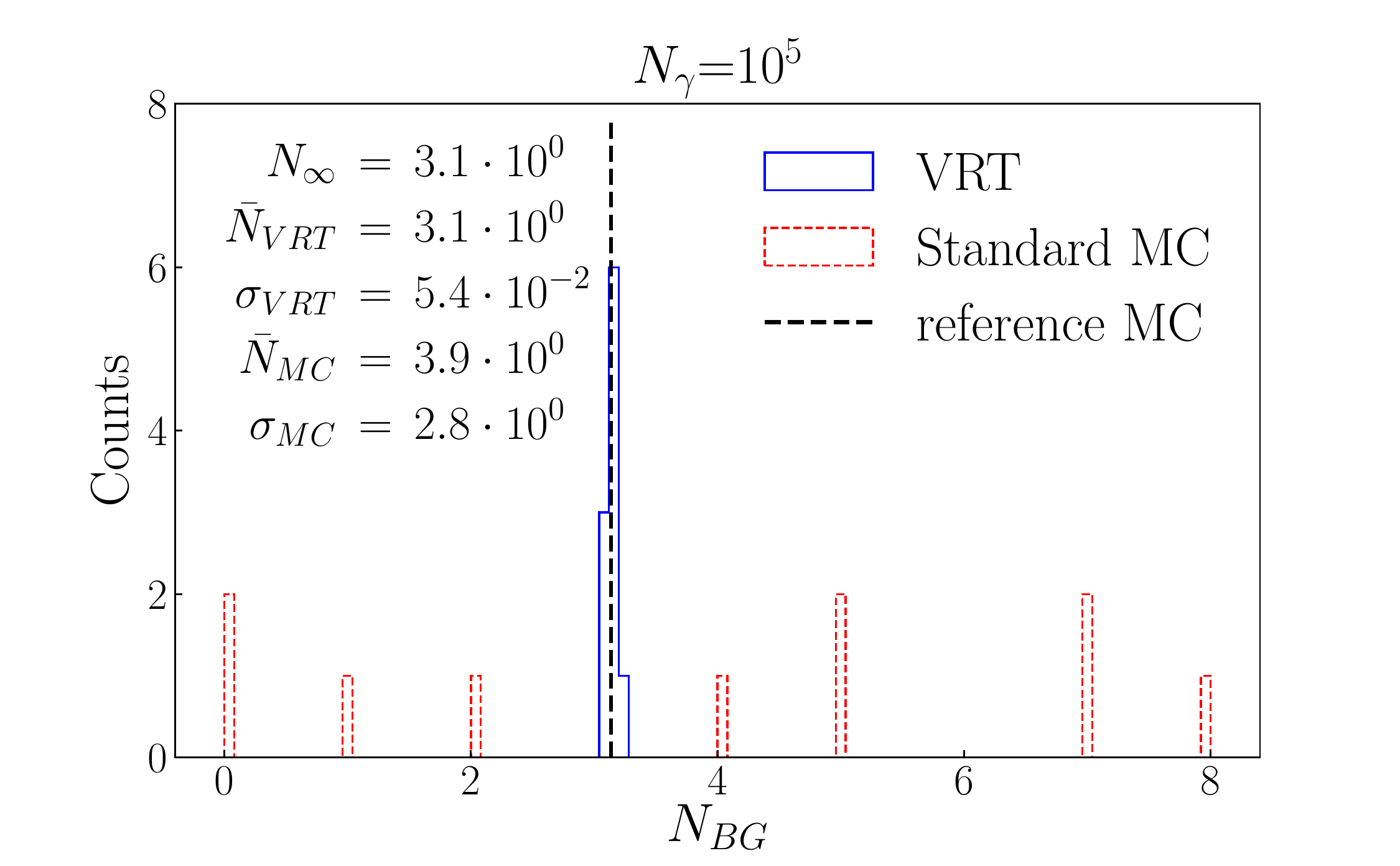}
        \vskip\baselineskip
            \includegraphics[width=0.48\textwidth]{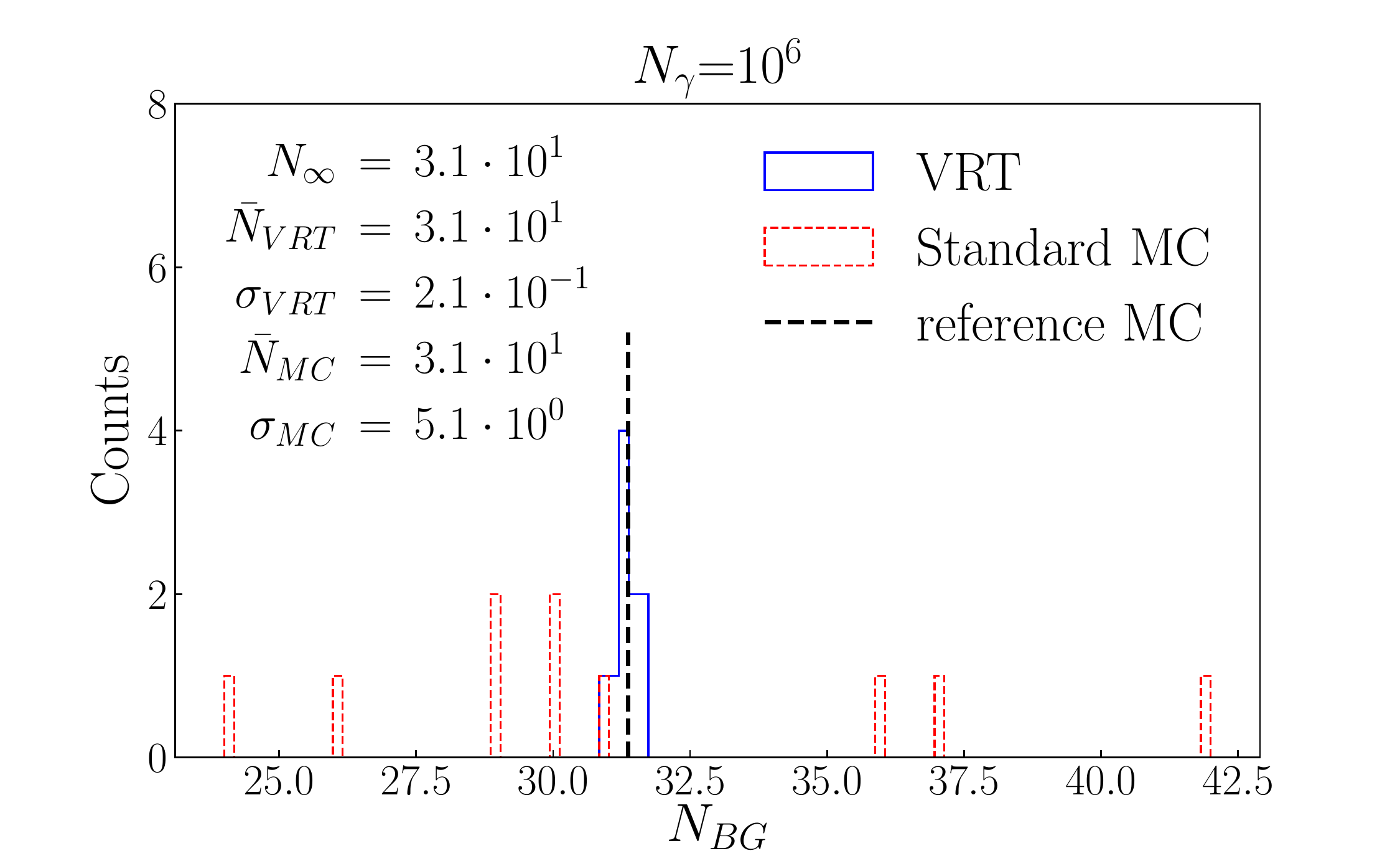}
        \hspace*{\fill}
            \includegraphics[width=0.48\textwidth]{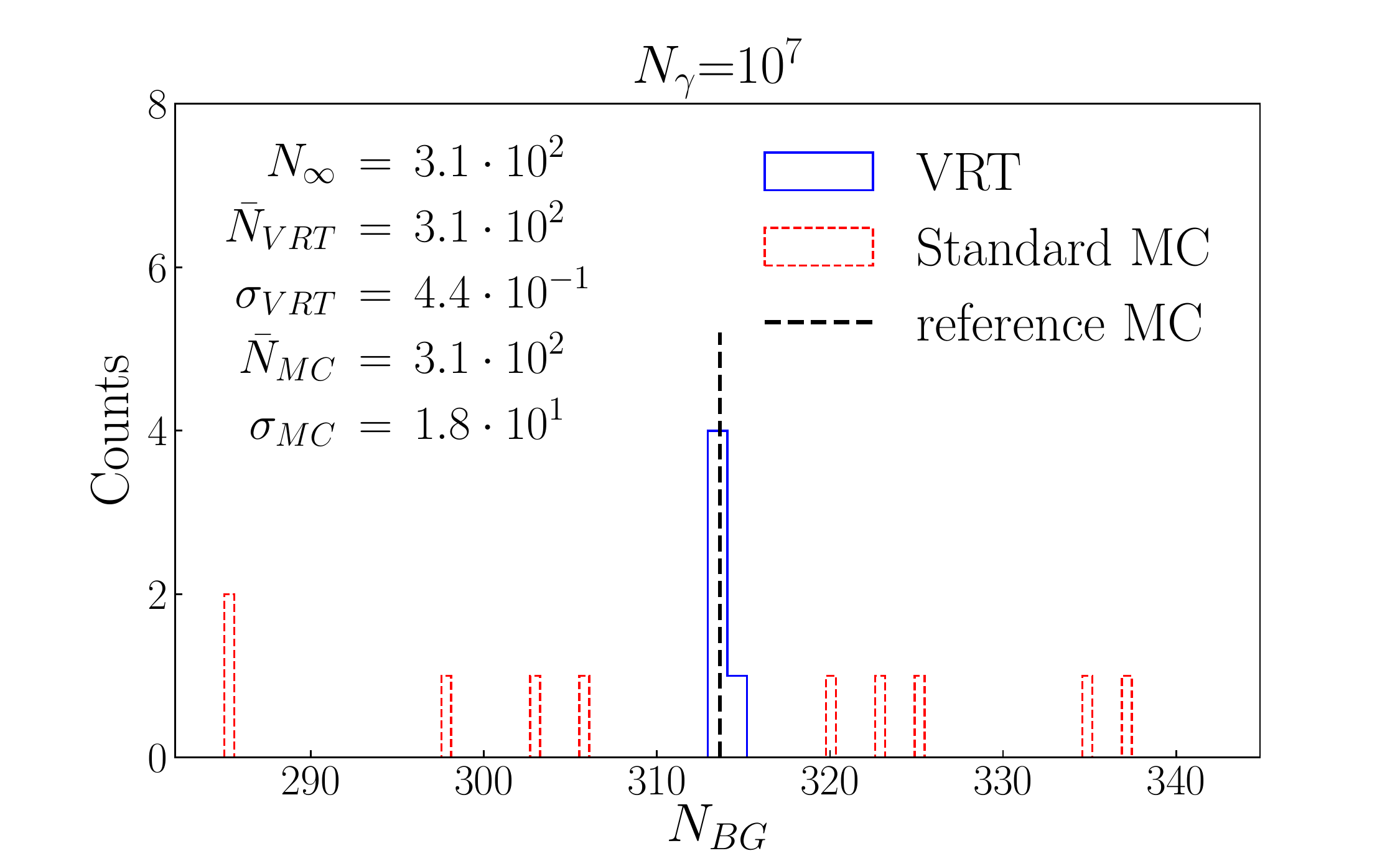}
        \caption{The number of gamma rays undergoing a single scatter inside the fiducial volume with an energy deposit of $<250\,$keV. Figures a-d, show the simulation results for an initial number of gamma rays of $N_{\gamma} =10^4$, $10^5$, $10^6$ and $10^7$, respectively. As a reference serves the expectation value $N_\infty$ of a high statistics simulation. In case of the accelerated method the expectation value of the predicted number of interactions in the ROI, $\bar{N}_{VRT}$, shows a much reduced variance, $\sigma_{VRT}$, with respect to the standard MC approach. }
        \label{fig:Total_weights}
    \end{figure*}
As described in Section \ref{sec:simulation}, our method gains its acceleration by accounting for some photon transport and physics processes by means of weights $p_i$ instead of determining those survival probabilities for an event to happen in the ROI by simulation. The increased sampling in the ROI leads to a more precise expectation value of the simulation, i.e., with a smaller standard deviation for a given number of simulated gammas.
This section quantifies the acceleration of our method compared to the standard MC method by determining the number of events for both simulations that yield the same statistical uncertainty on the estimated events in the ROI. This acceleration will depend on the detector geometry and energy of the gamma rays, but it is independent of computing power.

To quantify the acceleration we perform for both MC methods a set of 10 independent simulations with a certain number of initial gammas $N_\gamma$ at $1.5\,$MeV. The goal is to determine the expected number of single-scatter events $N_{BG}$ within a defined FV. For these simulations the detector geometry and FV were fixed as shown in Figure \ref{fig:setup} and an energy cut was set to $E_{max} = 250\,$keV. The results obtained for four different $N_\gamma$ are shown in Figure \ref{fig:Total_weights}.

In all cases we observe that the expectation values $N_{BG}$ obtained by the standard MC method for a certain $N_\gamma$ have a much larger spread than the results from the acceleration method. For both methods the means, $\bar{N}_{MC}$ and $\bar{N}_{VRT}$ respectively, converge to the expectation value of a high statistics reference simulation referred to as  $N_{\infty}$. However, the sample variance $\sigma_{MC}$ for the standard MC is much larger with respect to $\sigma_{VRT}$ in the accelerated simulation.\\

\begin{figure}
    \centering
    \includegraphics[width=0.5\textwidth]{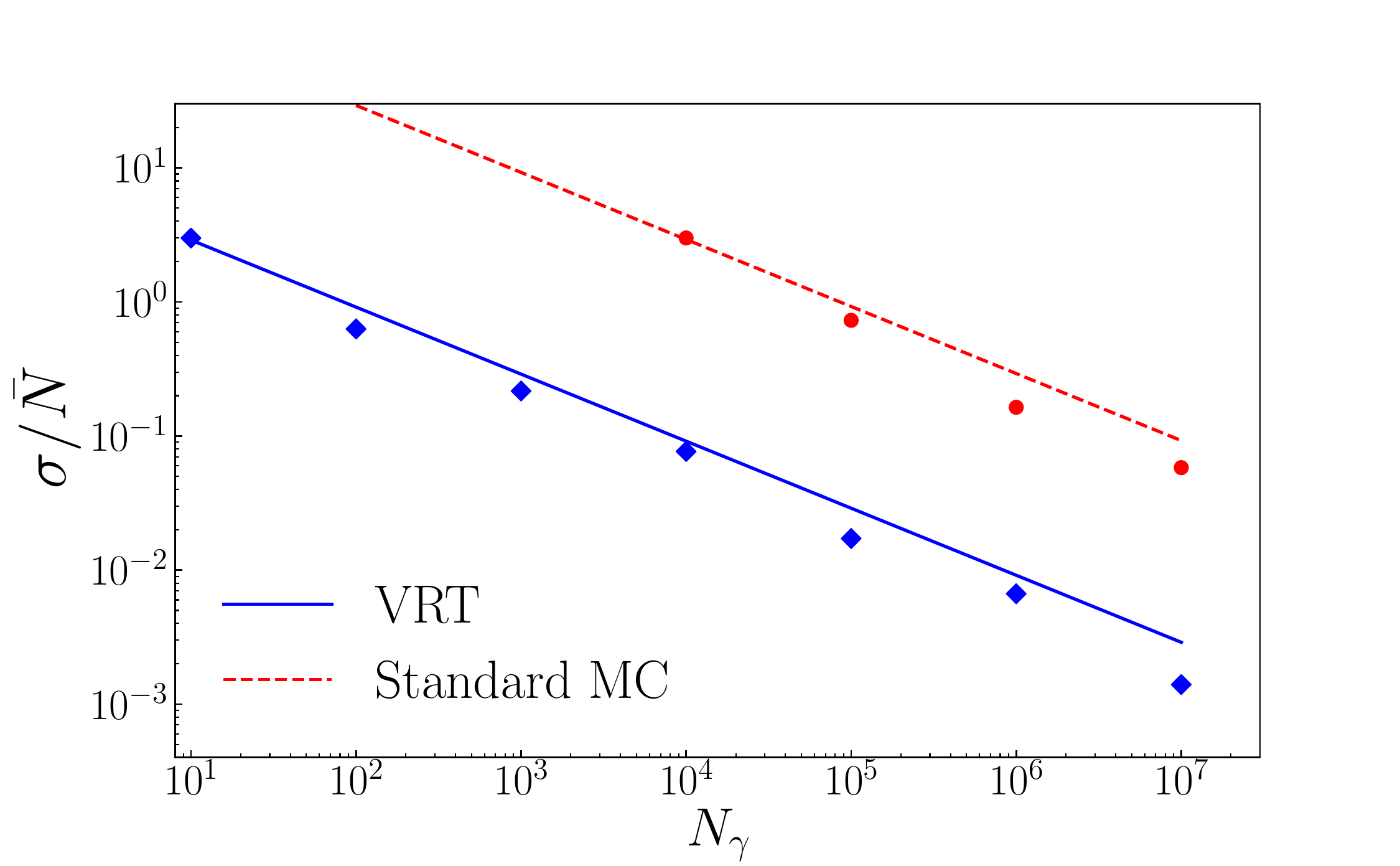}
    \caption{Evolution of the relative sample variance as a function of $N_\gamma$ for the standard MC (red) and the accelerated VRT MC (blue) simulation. The lines correspond to a fit of equation \ref{eq:fitfun} to the data points.}
    \label{fig:sigma_n}
\end{figure}

Figure \ref{fig:sigma_n} depicts the evolution of the relative uncertainty $\sigma / \bar{N}$ as a function of the number of simulated events for the standard MC and the accelerated VRT MC. For the simulations with $N_\gamma < 10^4$ no interaction happened in the ROI in the standard MC approach. 
The same relative error is achieved with a smaller number of simulated events for the accelerated MC, due to the more efficient sampling of the underlying probability density functions in the ROI. For both simulations the dependence of the relative statistical uncertainty on $N_{\gamma}$ is parameterized as
\begin{equation}
\label{eq:fitfun}
\sigma/\bar{N} = \frac{c}{\sqrt{N_{\gamma}}}    \quad .
\end{equation}
The fit model provides only a qualitative description of the data. Nevertheless we use the fit-parameters to estimate the achieved acceleration for a particular simulation geometry.
 The acceleration factor, $\alpha$, is defined as the ratio of the number of events
 using the accelerated MC, $N_{\gamma}^{\scriptscriptstyle{VRT}}$, and the standard MC $N_{\gamma}^{\scriptscriptstyle{MC}}$, when both methods achieve a certain relative statistical uncertainty, i.e.,
\begin{equation}
    \alpha = \frac{N_{\gamma}^{\scriptscriptstyle{VRT}}}{N_{\gamma}^{\scriptscriptstyle{MC}}} = \left(\frac{c^{\scriptscriptstyle{MC}}}{c^{\scriptscriptstyle{VRT}}}\right)^2 \quad ,
\end{equation}
with the fitted slopes $c^{\scriptscriptstyle{MC}}$ and $c^{\scriptscriptstyle{VRT}}$ from equation~\ref{eq:fitfun}. In our particular geometry we achieve an acceleration factor of $1800$ for single scatter events in the ROI. Due to the multidimensionality of a Monte Carlo simulations it is impossible to predict in advance what the acceleration factor will be. It strongly depends on the particulars of the chosen geometry and the energy of the gamma rays. The acceleration technique will generally benefit the simulation of extremely rare events due to the efficient sampling of the underlying pdf in the ROI.

\section{Discussion and Conclusion}

This paper presents the proof of principle for a variance reduction technique that can be used to estimate gamma ray backgrounds in large dark matter experiments. The evaluation of the method showed an acceleration factor $\alpha= 1800$ with respect to a standard non-weighted Monte Carlo simulation. The acceleration factor will depend on the geometry of the system, the size of the fiducial volume,  and the energy of the gamma rays. As a general rule
of thumb, the larger the detector volume, the larger $\alpha$ becomes, due to the fact that the
survival probability of the gamma rays decreases exponentially  with increased pathlength.
Similarly, the maximal allowed energy deposition has a strong impact on the achieved acceleration factor. If the cut-off is at low energy deposits, as in the signal-region for WIMP nuclear recoils, the  acceleration factor is large, as simulation of not contributing photo-electric absorption events is eliminated. Furthermore, we observed that $\alpha$ varied by a factor of $\sim 3$ over the energy range of 500~keV to 1.5~MeV, due to strong variation of the photo-electric and incoherent cross-sections.

One important aspect of simulations for WIMP detection experiments is the capability to simulate neutrons, since these may cause an irreducible background. For this proof of principle study we chose to simulate gamma rays only, since the physics processes relevant to their transport is relatively simple. The same acceleration method is applicable to neutron transport as well and could be implemented using e.g. GEANT4~\cite{Geant4:2003}. 


\begin{figure*}
    \centering
    \includegraphics[width=0.7\textwidth]{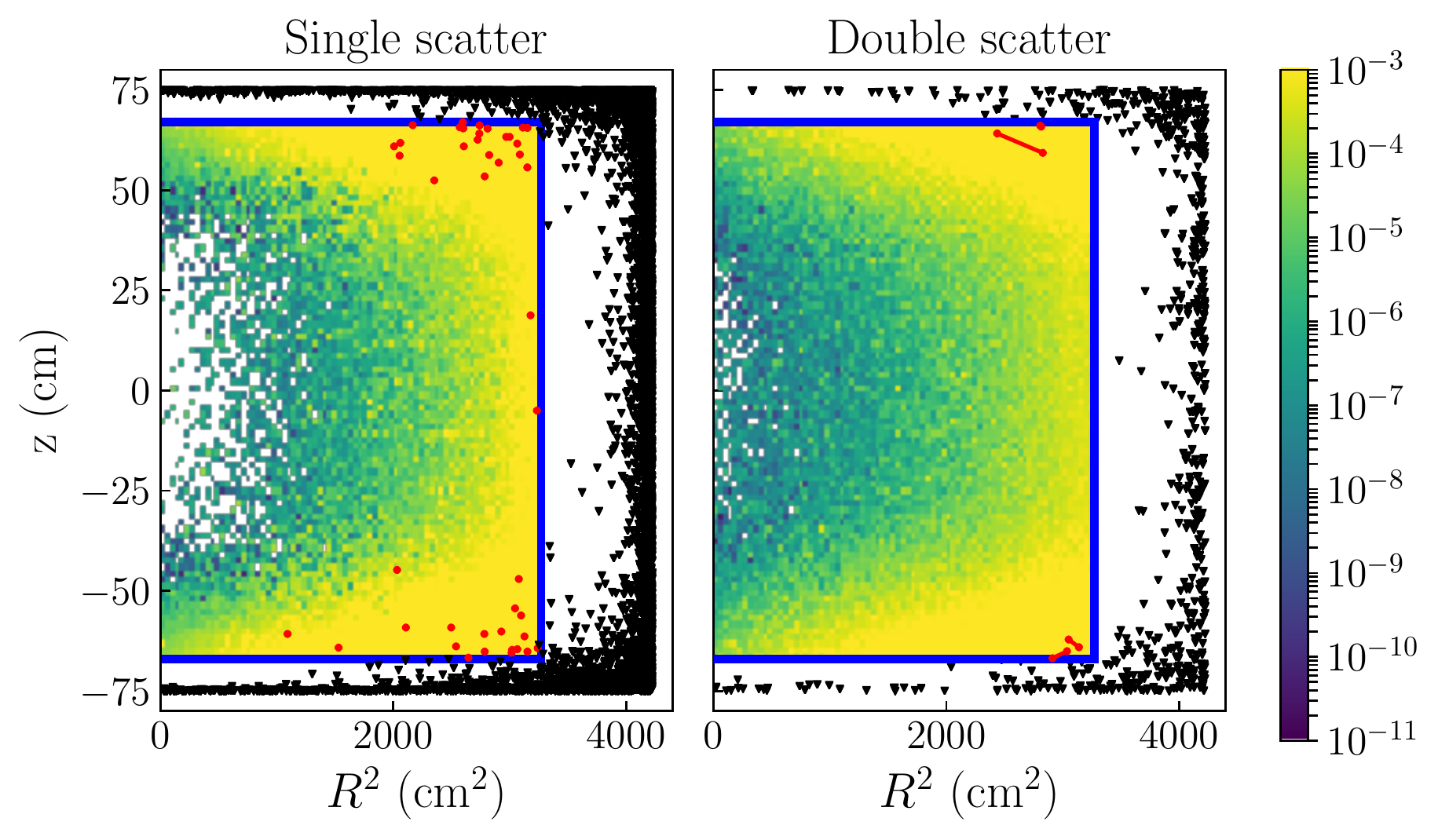}
    \caption{Spatial distribution of single scatter (left) and double scatter events (right) from a simulation of $10^6$ gamma rays of 1~MeV energy. The total energy deposit in the target is constraint below 250~keV. The dots represent the energy deposits from the standard Monte Carlo, while the colored distribution shows the energy deposits from the accelerated simulation inside the FV. For the double scatter events simulated with the standard MC the red lines are connecting the first and second scatter inside the FV. The standard MC gives sparse information inside the FV both for single scatters and for double scatters. The accelerated MC gives a more detailed - low variance - estimate of the spatial distribution of the background events.}
    \label{fig:double_scatter}
\end{figure*}

Closely related to the capability to simulate neutrons, is the ability to simulate multiple-scatter events. Neutrons, in sharp contrast to WIMPs, have a sizeable probability to undergo a double scatter, and it is therefore of crucial importance to acquire detailed knowledge of such events. Although we quantified the acceleration factor for single scatter events only, the simulation code is set-up to simulate multiple-scatters. Figure~\ref{fig:double_scatter} shows the spatial distribution of single and double scattered 1~MeV gamma rays for both the standard Monte Carlo and for the accelerated simulation. Both interactions were required to happen inside the fiducial volume, while the total energy deposit was required to be below 250~keV. For the standard Monte Carlo only very few events are accepted, resulting in a relatively high uncertainty on this background, whereas the accelerated Monte Carlo shows a smooth distribution of events. Besides offering a much more precise estimate of this background, this also allows to investigate the details of such events. For example, based on the accelerated simulation it is possible to clearly identify which detector components contribute significantly to the background, and take these results into consideration in
the detector design.

To summarize, a new variance reduction technique for photon transport was developed and validated. Much faster estimates of backgrounds can be obtained, and this method has the potential to accelerate simulations for neutron background estimates and multiple scatter events allowing their study in 
much greater detail. 

\bibliographystyle{unsrtnat}

\bibliography{bibliography.bib}

\begin{thebibliography}{15}
\providecommand{\natexlab}[1]{#1}
\providecommand{\url}[1]{\texttt{#1}}
\expandafter\ifx\csname urlstyle\endcsname\relax
  \providecommand{\doi}[1]{doi: #1}\else
  \providecommand{\doi}{doi: \begingroup \urlstyle{rm}\Url}\fi

\bibitem[Aprile et~al.(2018)]{Aprile:2018final}
E.~Aprile et~al.
\newblock {Dark Matter Search Results from a One Tonne x Year Exposure of
  XENON1T}.
\newblock \emph{Phys. Rev. Lett.}, 121:\penalty0 111302, 2018.

\bibitem[Akerib et~al.(2020{\natexlab{a}})]{LZ:2018}
D.~S. Akerib et~al.
\newblock Projected wimp sensitivity of the lux-zeplin dark matter experiment.
\newblock \emph{Phys. Rev. D}, 101:\penalty0 052002, Mar 2020{\natexlab{a}}.
\newblock \doi{10.1103/PhysRevD.101.052002}.
\newblock URL \url{https://link.aps.org/doi/10.1103/PhysRevD.101.052002}.

\bibitem[Anton et~al.(2019)]{EXO-200:2019}
G.~Anton et~al.
\newblock Search for neutrinoless double-$\ensuremath{\beta}$ decay with the
  complete exo-200 dataset.
\newblock \emph{Phys. Rev. Lett.}, 123:\penalty0 161802, Oct 2019.
\newblock \doi{10.1103/PhysRevLett.123.161802}.
\newblock URL \url{https://link.aps.org/doi/10.1103/PhysRevLett.123.161802}.

\bibitem[Aprile et~al.(2017)]{Aprile:2017screening}
E.~Aprile et~al.
\newblock Material radioassay and selection for the xenon1t dark matter
  experiment.
\newblock \emph{The European Physical Journal C}, 77\penalty0 (12):\penalty0
  890, 2017.
\newblock \doi{10.1140/epjc/s10052-017-5329-0}.
\newblock URL \url{https://doi.org/10.1140/epjc/s10052-017-5329-0}.

\bibitem[Akerib et~al.(2020{\natexlab{b}})]{Akerib:2020com}
D.S. Akerib et~al.
\newblock {The LUX-ZEPLIN (LZ) radioactivity and cleanliness control programs}.
\newblock 6 2020{\natexlab{b}}.

\bibitem[Aprile et~al.(2016)]{Aprile:2015uzo}
E.~Aprile et~al.
\newblock {Physics reach of the XENON1T dark matter experiment}.
\newblock \emph{JCAP}, 1604\penalty0 (04):\penalty0 027, 2016.
\newblock \doi{10.1088/1475-7516/2016/04/027}.

\bibitem[Akerib et~al.(2015)]{LUX:2015MC}
D.S. Akerib et~al.
\newblock Radiogenic and muon-induced backgrounds in the lux dark matter
  detector.
\newblock \emph{Astroparticle Physics}, 62:\penalty0 33 -- 46, 2015.
\newblock ISSN 0927-6505.
\newblock \doi{https://doi.org/10.1016/j.astropartphys.2014.07.009}.
\newblock URL
  \url{http://www.sciencedirect.com/science/article/pii/S0927650514001054}.

\bibitem[Aalbers et~al.(2016)]{Aalbers:2016DARWIN}
J.~Aalbers et~al.
\newblock {DARWIN}: towards the ultimate dark matter detector.
\newblock \emph{JCAP}, 2016\penalty0 (11):\penalty0 017--017, nov 2016.
\newblock \doi{10.1088/1475-7516/2016/11/017}.
\newblock URL \url{https://doi.org/10.1088%2F1475-7516%2F2016%2F11%2F017}.

\bibitem[Tokdar and Kass(2010)]{doi:10.1002/wics.56}
Surya~T. Tokdar and Robert~E. Kass.
\newblock Importance sampling: a review.
\newblock \emph{WIREs Computational Statistics}, 2\penalty0 (1):\penalty0
  54--60, 2010.
\newblock \doi{10.1002/wics.56}.
\newblock URL \url{https://onlinelibrary.wiley.com/doi/abs/10.1002/wics.56}.

\bibitem[NIS(2010)]{NIST}
Xcom photon cross sections database, 2010.
\newblock retrieved on April 1, 2019,
  \url{https://www.nist.gov/pml/xcom-photon-cross-sections-database}.

\bibitem[Aprile et~al.(2019)]{Aprile:2019bbb}
E.~Aprile et~al.
\newblock {XENON1T Dark Matter Data Analysis: Signal Reconstruction,
  Calibration and Event Selection}.
\newblock \emph{Phys. Rev. D}, 100\penalty0 (5):\penalty0 052014, 2019.
\newblock \doi{10.1103/PhysRevD.100.052014}.

\bibitem[Hubbell et~al.(1975)]{Hubbell:1975formfactors}
J.~H. Hubbell et~al.
\newblock Atomic form factors, incoherent scattering functions, and photon
  scattering cross sections.
\newblock \emph{Journal of Physical and Chemical Reference Data}, 4\penalty0
  (3):\penalty0 471--538, 1975.
\newblock \doi{10.1063/1.555523}.
\newblock URL \url{https://doi.org/10.1063/1.555523}.

\bibitem[Akerib et~al.(2020{\natexlab{c}})]{Akerib:2019fml}
D.S. Akerib et~al.
\newblock {The LUX-ZEPLIN (LZ) Experiment}.
\newblock \emph{Nucl. Instrum. Meth. A}, 953:\penalty0 163047,
  2020{\natexlab{c}}.
\newblock \doi{10.1016/j.nima.2019.163047}.

\bibitem[Aprile et~al.(2020)]{xenonnt}
E.~Aprile et~al.
\newblock {Projected WIMP Sensitivity of the XENONnT Dark Matter Experiment}.
\newblock \emph{submitted to JCAP}, 2020.

\bibitem[Agostinelli et~al.(2003)]{Geant4:2003}
S.~Agostinelli et~al.
\newblock Geant4 - a simulation toolkit.
\newblock \emph{Nuclear Instruments and Methods in Physics Research Section A:
  Accelerators, Spectrometers, Detectors and Associated Equipment},
  506\penalty0 (3):\penalty0 250 -- 303, 2003.
\newblock ISSN 0168-9002.
\newblock \doi{https://doi.org/10.1016/S0168-9002(03)01368-8}.
\newblock URL
  \url{http://www.sciencedirect.com/science/article/pii/S0168900203013688}.

\end{thebibliography}

\end{document}